\documentclass[twocolumn]{svjour3}

\usepackage{amsmath,amsfonts,bm,graphicx,amssymb,color,mathrsfs,mathptmx,units}

\newcommand{\llangle}{\langle\!\langle}
\newcommand{\rrangle}{\rangle\!\rangle}
\newcommand{\GBequal}{\underset{\mathrm{binomial}}{\overset{\mathrm{generalized}}{=}}}

\journalname{Journal of Computational Electronics}

\begin{document}

\title{Time-dependent factorial cumulants in interacting nano-scale systems}

\author{Dania Kambly         \and
        Christian Flindt
}

\institute{D. Kambly \and C. Flindt \at D\'epartement de Physique Th\'eorique, Universit\'e de Gen\`eve, CH-1211 Gen\`eve, Switzerland \\ \email{dania.kambly@unige.ch \\ christian.flindt@unige.ch}}

\date{Received: date / Accepted: date}

\maketitle

\begin{abstract}
We discuss time-dependent factorial cumulants in interacting nano-scale systems. Recent theoretical work has shown that the full counting statistics of non-interacting electrons in a two-terminal conductor is always generalized binomial and the zeros of the generating function are consequently real and negative. However, as interactions are introduced in the transport, the zeros of the generating function may become complex. This has measurable consequences: With the zeros of the generating function moving away from the real-axis, the high-order factorial cumulants of the transport become oscillatory functions of time. Here we demonstrate this phenomenon on a model of charge transport through coherently coupled quantum dots attached to voltage-biased electrodes. Without interactions, the factorial cumulants are monotonic functions of the observation time. In contrast, as interactions are introduced, the factorial cumulants oscillate strongly as functions of time. We comment on possible measurements of oscillating factorial cumulants and outline several avenues for further investigations.

\end{abstract}

% 02.50.Ey Stochastic processes
% 72.70.+m Noise processes and phenomena
% 73.23.Hk Coulomb blockade; single-electron tunneling

\section{Introduction}
\label{sec:introduction}

The full counting statistics (FCS) of charge transfers in sub-micron electrical conductors has become an active field of research~\cite{Levitov1993,Blanter2000,Nazarov2003}. Initially, investigations of FCS were primarily of theoretical interest, but several experiments~\cite{Reulet2003,Lu2003,Bomze2005,Bylander2005,Fujisawa2006,Gustavsson2006,Fricke2007,Sukhorukov2007,Timofeev2007,Gershon2008,Gustavsson2009,Gabelli2009,Flindt2009,Fricke2010a,Fricke2010b,Ubbelohde2012} have now clearly demonstrated that measurements of FCS are achievable and much progress has been made: Non-Gaussian voltage and current fluctuations have been measured in tunnel junctions~\cite{Reulet2003,Bomze2005,Timofeev2007} and quantum point contacts~\cite{Gershon2008}, and the fourth and fifth current cumulants have been detected in an avalanche diode~\cite{Gabelli2009}.  Additionally, real-time electron detection techniques~\cite{Lu2003,Bylander2005} have paved the way for measurements of the FCS of charge transport in single~\cite{Gustavsson2006,Fricke2007,Sukhorukov2007,Flindt2009,Fricke2010a,Fricke2010b,Ubbelohde2012} and double quantum dots~\cite{Fujisawa2006,Gustavsson2009}. Following the initial measurements of the third cumulant of transport through quantum dots~\cite{Fujisawa2006,Gustavsson2006}, a series of experiments have addressed the conditional FCS~\cite{Sukhorukov2007}, the transient high-order cumulants~\cite{Flindt2009,Fricke2010a,Fricke2010b}, and the finite-frequency FCS~\cite{Ubbelohde2012} in quantum dot systems.  The works on transient high-order cumulants showed that high-order cumulants generically oscillate as functions of basically any system parameter as well as the observation time~\cite{Flindt2009}.

Investigations of FCS are motivated by the expectation that more information about the fundamental transport mechanisms can be extracted from the full statistical distribution of transferred charges than from the mean current and shot noise only~\cite{Levitov1993,Blanter2000,Nazarov2003}. However, the fact that high-order cumulants generically oscillate makes it less clear exactly what information the high-order cumulants contain? In a recent work~\cite{Kambly2011}, we have been drawing attention to the use of \emph{factorial} cumulants to characterize the FCS of charge transport in nano-scale electrical conductors. So far, factorial cumulants have only received limited attention in mesoscopic physics (but see Refs.~\cite{Beenakker2004,Song2011,Song2012,Komijani2013}). However, as we have shown, the factorial cumulants never oscillate (unlike the ordinary cumulants) for non-interacting two-terminal scattering problems. This result is based on the recent finding that the FCS for non-interacting electrons in a two-terminal scattering setup is always generalized binomial and the zeros of the generating function for the FCS consequently are real and negative~\cite{Abanov2008,Abanov2009}; see Ref.~\cite{Kambly2009} for a discussion of multi-terminal conductors. In contrast, as interactions are introduced in the transport, the zeros of the generating function may become complex and the factorial cumulants start to oscillate~\cite{Kambly2011}. This indicates that factorial cumulants may be useful to detect interactions among charges passing through a nano-sized electrical conductor. As such we address the fundamental question concerning FCS, namely what we can learn about a physical system by measuring the transport statistics beyond the mean current and the noise.

The purpose of this work is to illustrate these ideas on a model of transport through coherently coupled quantum dots. In previous work~\cite{Kambly2011,Kambly2012}, we considered systems described by classical master equations. We now turn to a situation, where the quantum coherent coupling between two parts of the conductor is important. The system we consider is a double quantum dot (DQD) attached to external source and drain electrodes. We employ a generalized master equation (GME) approach which allows us to treat strong coupling to the leads together with the coherent evolution of electrons inside the DQD. We treat two cases of particular interest: In the non-interacting regime, the DQD can accommodate zero, one, or two electrons at a time, without additional charging energy required for the second electron. We show that the factorial cumulants in this case do not oscillate as functions of the observation time and from the high-order factorial cumulants we extract the zeros of the generating function which are real and negative. Next, we consider the strongly interacting case, where double-occupation of the DQD is excluded. In this case, the time-dependent factorial cumulants oscillate -- a clear signature of interactions in the transport -- and the zeros of the generating function are complex.

In the interacting case, we find that the Fano factor $F$, i.~e.~the ratio of the shot noise over the mean current, may either be super-Poissonian ($F>1$) or sub-Poissonian~($F<1$). Super-Poissonian noise is typically taken as a signature of interactions in the transport~\cite{Blanter2000}, while no clear conclusion can be drawn from a sub-Poissonian Fano factor. Interestingly, we find that the factorial cumulants may oscillate in both situations, showing that the factorial cumulants can provide a clear signature of interactions even when the current fluctuations are sub-Poissonian. We conclude our theoretical investigations of factorial cumulants by examining the influence of dephasing of electrons passing through the DQD~\cite{Kiesslich2006,Braggio2009}, for instance due to a nearby charge detector.

The paper is organized as follows: in Sec.~\ref{sec:counting_statistics} we introduce the essential terminology  used in FCS and provide the basic definitions with a special emphasis on \emph{factorial} cumulants and the concept of generalized binomial statistics. In Sec.~\ref{sec:highcumu} we then turn to the asymptotic behavior of high-order cumulants (both ordinary and factorial cumulants) and show why the high-order factorial cumulants do not oscillate for transport of non-interacting electrons in a two-terminal conductor. In Sec.~\ref{sec:DQD} we introduce a model of electron transport through a DQD described by a Markovian GME, while Sec.~\ref{sec:calculations} is devoted to the details of our calculations of time-dependent factorial cumulants. In Sec.~\ref{sec:results} we demonstrate how interactions on the DQD give rise to clear oscillations of the high-order factorial cumulants with the zeros of the generating function moving away from the negative real-axis and into the complex plane. Finally, Section~\ref{sec:conclusions} is dedicated to a summary of the work as well as our concluding remarks.

\section{Full counting statistics \& factorial cumulants}
\label{sec:counting_statistics}

Full counting statistics concerns the quantum statistical process of electron transport in mesoscopic conductors~\cite{Levitov1993,Blanter2000,Nazarov2003,Kambly2011,Abanov2008,Abanov2009,Kambly2009,Kambly2012,Kiesslich2006,Braggio2009,Ivanov1993,Levitov1996,Nazarov2002,Pilgram2003,Nagaev2004,Bagrets2003,Emary2007,Marcos2010,Braggio2006,Flindt2008,Flindt2010,Marcos2011,Emary2011,Vanevic2007,Vanevic2008,Hassler2008,Esposito2009,Foerster2008,Sanchez2010}. The full counting statistics (FCS) is the probability $P(n,t)$ that $n$ electrons have traversed a conductor during a time span $[0,t]$ of duration $t$. The information contained in the probability distribution may equally well be encoded in the generating function (GF) defined as
\begin{equation}\label{Eq:GF}
    \mathcal{G}(z,t)=\sum_{n}P(n,t)z^n.
\end{equation}
The normalization condition for the probabilities, $\sum_nP(n,t)=1$, implies for the GF that $\mathcal{G}(z=1,t)=1$. Important information about the charge transport can be obtained from the GF: If the transport process consists of several independent sub-processes, the GF factors into a product of the GFs corresponding to each of these sub-processes, similarly to how the partition function in statistical mechanics may be written as a product of the partition functions for each independent sub-system. Elementary transport processes can thus be identified by factorizing the GF. In the case of transport of noninteracting electrons through a two-terminal conductor, Abanov and Ivanov have shown recently that the GF can be factorized into single-particle events of binomial form \cite{Abanov2008,Abanov2009} (also see \cite{Vanevic2007,Vanevic2008}). Such distributions have been dubbed generalized binomial statistics \cite{Hassler2008,Abanov2008,Abanov2009}, which will be of central importance in this work.

Several useful statistical functions and quantities can be obtained from the GF. First, we can define a moment generating function (MGF)
\begin{equation}\label{Eq:MGF}
 \mathcal{M}(z,t)=\mathcal{G}(e^z,t),
\end{equation}
which generates the statistical moments of $n$ by differentiation with respect to the counting field $z$ at $z=0$
\begin{equation}\label{Eq:moment}
\langle n^m\rangle(t)=\partial^m_z\mathcal{M}(z,t)|_{z\rightarrow 0}=\sum_n n^m P(n,t).
\end{equation}
The MGF of a transport process composed of several independent processes factors into a product of the corresponding MGFs.  However, the moments of the full process are not related to the moments of the individual sub-processes in a simple way. This motivates the definition of cumulants, also known as irreducible moments. The cumulant generating function (CGF) is defined as the logarithm of the MGF
\begin{equation}\label{Eq:MGF}
\mathcal{S}(z,t)=\log[\mathcal{M}(z,t)]=\log[\mathcal{G}(e^z,t)],
\end{equation}
which again delivers the cumulants of $n$ by differentiation with respect to $z$ at $z=0$:
\begin{equation}\label{Eq:cumu}
\langle\!\langle n^m \rangle\!\rangle(t)=\partial^m_z\mathcal{S}(z,t)|_{z\rightarrow 0}.
\end{equation}
The first cumulant is the mean of $n$, $\langle\!\langle n \rangle\!\rangle= \langle n \rangle$,  the second cumulant is the variance, $\langle\!\langle n \rangle\!\rangle= \langle n^2 \rangle-\langle n \rangle^2$, and the third is the skewness, $\langle\!\langle n^3 \rangle\!\rangle= \langle (n-\langle n \rangle)^3 \rangle$. It is easy to show that the cumulants of a transport process are simply the sum of the cumulants corresponding to each independent sub-process. Moreover, for a Gauss distribution only the first and second cumulants are non-zero, while all higher cumulants vanish. In this respect, one may use cumulants of a distribution as a measure of (non-)gaussianity.

The conventional moments and cumulants, as defined above, have been investigated intensively in the field of FCS \cite{Nazarov2003}. The zero-frequency cumulants of the current are given by the long-time limit of the cumulants of $n$ as
\begin{equation}\label{Eq:current_cumulants}
    \langle\!\langle I^m \rangle\!\rangle\equiv\lim_{t\rightarrow\infty}\frac{\langle\!\langle n^m \rangle\!\rangle}{t}.
\end{equation}
The current is treated as a \emph{continuous} variable and continuous variables are typically characterized by their cumulants. However, another interesting class of statistical quantities exists, which has received much less attention in FCS. These are the factorial moments and the factorial cumulants, which are mostly discussed in the context of \emph{discrete} variables\cite{Kendall1977,Johnson2005}. The number of counted electrons $n$ is obviously a discrete variable, and it is natural to ask if the current cumulants, as defined in Eq.~(\ref{Eq:current_cumulants}), carry signatures of this discreteness?

The factorial moments are again generated by a factorial MGF, which can be defined based on the GF in Eq.~(\ref{Eq:GF}). The factorial MGF is defined as
\begin{equation}\label{Eq:FMGF}
    \mathcal{M}_F(z,t)=\mathcal{G}(z+1,t)
\end{equation}
and the corresponding factorial moments read
\begin{equation}\label{Eq:FactMom}
    \langle n^m \rangle_F(t)\equiv\partial^m_z \mathcal{M}_F(z,t)|_{z\rightarrow 0}=\langle n(n-1)\dotsm (n-m+1)\rangle
\end{equation}
in terms of the ordinary moments. In analogy with the conventional CGF, the factorial CGF is defined as
\begin{equation}\label{Eq:FCGF}
    \mathcal{S}_F(z,t)=\log[\mathcal{M}_F(z,t)]=\log[\mathcal{G}(z+1,t)]
\end{equation}
and the corresponding factorial cumulants read
\begin{equation}\label{Eq:FactCum}
    \langle\!\langle n^m \rangle\!\rangle_F(t)\equiv\partial^m_z \mathcal{S}_F(z,t)|_{z\rightarrow 0}=\langle\!\langle n(n-1)\dotsm (n-m+1)\rangle\!\rangle.
\end{equation}
As mentioned above, factorial moments and factorial cumulants are of particular interest when considering probability distributions of discrete variables. For example, for a Poisson process with rate $\Gamma$, which is the physical limit of rare events, the FCS is well-known and reads
\begin{equation}
P(n,t)=\frac{(\Gamma t)^n}{n!}e^{-\Gamma t}.
\end{equation}
The corresponding GF then becomes
\begin{equation}
\mathcal{G}(z,t)=e^{\Gamma t(z-1)},\,\,\,\,\, \mathrm{(Poisson\, process)}
\end{equation}
from which it is easy to show that the cumulants are
\begin{equation}
\langle\!\langle n^m \rangle\!\rangle(t) = \Gamma t,\,\,\,\,\, \mathrm{(Poisson\, process)}
\end{equation}
for all $m=1,2,\ldots$. In contrast, the first factorial cumulant reads
\begin{equation}
\langle\!\langle n\rangle\!\rangle_F(t) = \Gamma t, \,\,\,\,\, \mathrm{(Poisson\, process)},
\end{equation}
while all higher factorial cumulants are zero
\begin{equation}
\langle\!\langle n^m\rangle\!\rangle_F(t)=0,\,\, m>1 \,\,\,\,\, \mathrm{(Poisson\, process)}.
\end{equation}
Thus, similarly to how ordinary cumulants are useful as measures of (non-)gaussianity, we may use factorial cumulants to characterize deviations of a distribution from Poisson statistics. It is also clear that a factorial cumulant of a given order is the sum of the factorial cumulants of all independent sub-processes, in the same way as for the cumulants.

Throughout this work we will rely on an important result by Abanov and Ivanov, who have shown that the FCS of non-interacting electrons in a two-terminal scattering problem is always generalized binomial \cite{Abanov2008,Abanov2009}. In this case, the GF takes the special form
\begin{equation}
\mathcal{G}(z,t)\GBequal z^{-Q}\prod_{i}(1-p_i+p_iz),
\label{Eq:GB_GF}
\end{equation}
where the (time-dependent) $p_i$'s are real with $0\leq p_i\leq 1$. Each factor in the product can be interpreted as a single binomial charge transfer event occurring with probability $p_i$. The factor in front, $z^{-Q}$, corresponds to a deterministic background charge transfer
\begin{equation}
Q=\sum_ip_i-\langle n\rangle\geq 0
\end{equation}
opposite to the positive direction of the mean current. For uni-directional transport, $\sum_ip_i=\langle n\rangle$ and $Q=0$, whereas $Q$ is non-zero for bi-directional transport due to thermal fluctuations for example. For uni-directional transport, the (time-dependent) Fano factor then reads
\begin{equation}\label{Eq:FanoFactor}
    F(t)\equiv\frac{\langle\!\langle n^2 \rangle\!\rangle(t)}{\langle n \rangle(t)}\GBequal=\frac{\sum_{i}p_i(1-p_i)}{\sum_{i} p_i},
\end{equation}
which is always smaller than unity, corresponding to a Poisson process.  Following this reasoning, a super-poissionian Fano factor, $F>1$, can be taken as a sign of interactions. Super-Poissonian noise was recently measured in an experiment on transport of interacting electrons through a double quantum dot \cite{Kiesslich2007}. Still, the noise may also be sub-Poissonian, $F<1$, in the presence of interactions.

\section{High (factorial) cumulants}
\label{sec:highcumu}

In this Section we are interested in the generic behavior of high-order (factorial) cumulants. To this end, we first discuss an approximation of high derivatives \cite{Dingle1973,Berry2005} and then show how these ideas can be applied in the context of FCS.

In the following, we consider a generic (factorial) CGF $\mathcal{S}_{(F)}$ and assume that it has a number of singularities $z_j$ in the complex plane. Close to each of these singularities, we may approximate the (factorial) CGF as
\begin{equation}
\label{eq:sing_approx}
\mathcal{S}_{(F)}(z,t)\simeq \frac{A_j}{\left(z_j-z\right)^{\mu_j}},\,\,\,\,{\rm for}\,\,\,\,\, z\simeq z_j
\end{equation}
for some constants  $A_j$ and $\mu_j$. Here, the constant $\mu_j$ is determined by the nature of the singularity, for instance $\mu_j=-\nicefrac{1}{2}$ corresponds to a square-root branch point, while an integer value of $\mu_j$ would correspond to the order of a pole at $z=z_j$. Using the first Darboux approximation \cite{Dingle1973,Berry2005,Flindt2009}, we may now evaluate the (factorial) cumulant of order $m$ by differentiating the expression in Eq.~(\ref{eq:sing_approx}) $m$ times with respect to $z$ at $z=0$ and sum over the contributions from all singularities as
\begin{equation}\label{Eq:DarbouxApprox}
    \langle\!\langle n^m \rangle\!\rangle_{(F)}\simeq\sum_j\frac{A_j B_{m,\mu_j}}{\left|z_j\right|^{m+\mu_j}}e^{-i\left(m+\mu_j\right)\arg z_j}.
\end{equation}
Here we have introduced the polar notation $z_j= |z_j|e^{i\arg z_j}$
together with the factors
\begin{equation}
B_{m,\mu_j}\equiv \mu_j(\mu_j+1)\dotsm(\mu_j+m-1).
\end{equation}

Equation (\ref{Eq:DarbouxApprox}) is particularly useful if the sum can be reduced to only a few terms. For high orders, the singularities closest to $z=0$ dominate the sum, which leads to a considerable simplification. For example, if a single complex conjugate pair of singularities, $z_0=|z_0|^{i\arg[z_0]}$ and $z_0^*=|z_0|^{-i\arg[z_0]}$, are closest to $z=0$, the high-order (factorial) cumulants can be approximated as
\begin{equation}\label{Eq:approxFC_one_pair}
    \langle\!\langle n^m \rangle\!\rangle_{(F)}\simeq\frac{2|A_0|B_{m,\mu_0}}{\left|z_0\right|^{m+\mu_0}}\cos\left[\left(m+\mu_0\right)\arg z_0-\arg A_0\right].
\end{equation}
This result shows that the absolute value of the (factorial) cumulants generically grows factorially with the cumulant order $m$, due to the factors $B_{m,\mu_0}$, and that they tend to oscillate as a function of \emph{any} parameter, including time $t$, that changes $\arg z_0$. Such universal oscillations have been observed experimentally in electron transport through a quantum dot \cite{Flindt2009,Fricke2010a,Fricke2010b}.

In contrast, in the particular situation, where there is just a single dominant singularity $z_0$ on the real-axis, the high-order (factorial) cumulants can be approximated as
\begin{equation}\label{Eq:approxFC_real_sng}
    \langle\!\langle n^m \rangle\!\rangle_{(F)}\simeq(-1)^{m+\mu_0}\frac{A_0B_{m,\mu_0}}{\left|z_0\right|^{m+\mu_0}}.
\end{equation}
In this case, the factorial growth with the order persists, but no oscillations are expected as long as the dominant singularity $z_0$ stays on the real-axis.

Let us now consider non-interacting electrons in a two-terminal conductor. As shown by Abanov and Ivanov~\cite{Abanov2008,Abanov2009}, the statistics is generalized binomial in this situation and the GF takes on the form given by Eq.~(\ref{Eq:GB_GF}). The corresponding cumulants are complicated functions of the probabilities $p_i$. In contrast, the factorial cumulants are simply
\begin{equation}
\langle\!\langle n^m\rangle\!\rangle_F \GBequal (-1)^{m-1}(m-1)!\left[\sum_i p_i^m-Q\right].
\label{Eq:GB_FCs}
\end{equation}
For uni-directional transport ($Q=0$), the largest probability $p_\mathrm{max}$ will dominate the high factorial cumulants, which can be approximated as
\begin{equation}
\langle\!\langle n^m\rangle\!\rangle_F \simeq (-1)^{m-1}(m-1)! p_\mathrm{max}^m.
\label{eq:largeFacGB}
\end{equation}
This expression can also be understood from Eq.~(\ref{Eq:DarbouxApprox}) by noting that the factorial CGF has logarithmic singularities at values of the counting field $z$, where the factorial MGF is zero. Combining Eqs.~(\ref{Eq:FMGF}) and (\ref{Eq:GB_GF}), we easily see that the factorial MGF has zeros at $z_j=-1/p_j\leq-1$. Moreover, the zero corresponding to the largest probability $p_\mathrm{max}$ is closest to $z=0$ and will dominate the high factorial cumulants as seen in Eq.~(\ref{eq:largeFacGB}).

We have seen above that high-order cumulants tend to oscillate as functions of basically any parameter, with or without interactions. In contrast, as our analysis also shows, the high-order factorial cumulants never oscillate for non-interacting electrons in a two-terminal scattering problem. This behavior can be traced back to the factorization of the GF in Eq.~(\ref{Eq:GB_GF}), which implies that the singularities of the factorial CGF are always real and negative. This makes factorial cumulants promising candidates for the detection of interactions in FCS. In particular, oscillating factorial cumulants must be due to interactions. In our previous work~\cite{Kambly2011}, we employed these ideas to incoherent electron transport through a single quantum dot. We showed how interactions may cause the singularities of the factorial CGF to move away from the real-axis and into the complex plane, making the high-order factorial cumulants oscillate.

In the following we apply these ideas to electron transport through a DQD, where the electrons may oscillate quantum coherently between the two quantum dots. In contrast to our previous work~\cite{Kambly2011}, we consider not only the time-dependent factorial cumulants of the transferred charge, but also the zero-frequency factorial cumulants of the current.

\section{Coulomb blockade quantum dots}
\label{sec:DQD}

We consider two-terminal nano-scale conductors connected to source and drain electrodes. Charge transport is described using a generalized master equation (GME) for the reduced density matrix $\hat\rho$ of the conductor obtained by tracing out the electronic leads. The GME accounts for the coherent evolution of charges inside the conductor as well as the transfer of electrons between the conductor and the leads. To evaluate the FCS, it is convenient to unravel the reduced density matrix with respect to the number of electrons $n$ that have been collected in the drain electrode during the time span $[0,t]$ \cite{Plenio1998,Makhlin2001}. With this $n$-resolved density matrix $\hat\rho(n,t)$ at hand, the FCS is obtained by tracing over the states of the conductor
\begin{equation}\label{Eq:ProbabilityDistr}
    P(n,t)=\mathrm{Tr}\left[\hat\rho(n,t)\right].
\end{equation}
Similarly, we recover the original reduced density matrix by summing over $n$, i.~e.
\begin{equation}
\hat\rho(t)=\sum_n\hat\rho(n,t).
\end{equation}
From these definitions, the GF reads
\begin{equation}
    \mathcal{G}(z,t)=\sum_n\mathrm{Tr}\left[\hat\rho(n,t)\right]z^n=\mathrm{Tr}\left[\hat\rho(z,t)\right],
\end{equation}
where we have introduced the $z$-dependent reduced density matrix
\begin{equation}
\label{eq:zdepRDM}
    \hat\rho(z,t)=\sum_n\hat\rho(n,t)z^n.
\end{equation}

\begin{figure}
\begin{center}
\includegraphics[width=0.4\textwidth, trim = 0 0 0 0, clip]{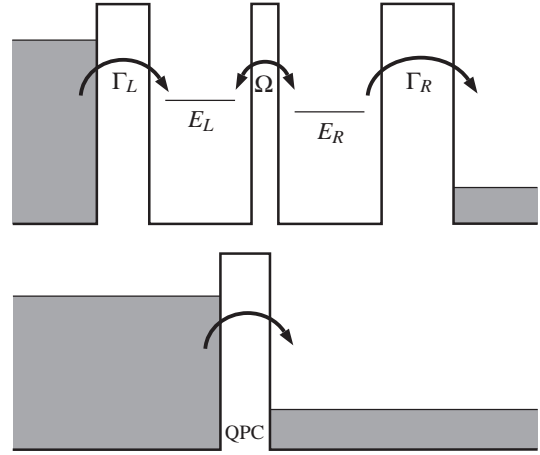}
\caption{Double quantum dot (DQD) coupled to a QPC charge detector. The upper panel shows the DQD. The tunnel coupling between the quantum dots is denoted as $\Omega$ and $E_L$ and $E_R$ are the single-particle levels of the two quantum dots with detuning $\epsilon\equiv E_R-E_L$. The tunneling rate from (to) the right (left) quantum dot to (from) the right (left) reservoir dot is denoted as $\Gamma_{R(L)}$.  The lower panel shows the QPC charge detector, which measures the charge occupation of the DQD. The QPC may couple asymmetrically to the DQD such that the charge occupation of the individual quantum dots can be resolved.}
\label{Fig:DQD}
\end{center}
\end{figure}

The particular conductor we now discuss consists of two quantum dots in series attached to source and drain electrodes. A schematic of the system is shown in Fig.~\ref{Fig:DQD}. The inter-dot Coulomb interaction can be tuned, so that both regimes of noninteracting and interacting electrons can be realized and compared. Disregarding the electronic spin degree of freedom (for example due to a strong magnetic field), the double quantum dot can be occupied by either zero, one, or two (additional) electrons. Experimentally, the charge on the quantum dots can be measured using a nearby quantum point contact (QPC), whose conductance is sensitive to the occupations of the individual quantum dots~\cite{Gustavsson2006,Gustavsson2009}. This charge detection protocol makes it possible to deduce the number of electrons that have passed through the DQD in a given time span. If the QPC is not sensitive to the charge occupations of the individual quantum dots, but only to the total charge, the measurement is not expected to destroy the coherent oscillations between the quantum dots. On the other hand, if the QPC measures the charge states of the individual quantum dots, it introduces decoherence in the dynamics of electrons inside the DQD~\cite{Gurvitz1996,Gurvitz1997}.

The full many-body Hamiltonian of our system reads
\begin{equation}\label{Eq:Hamiltonian}
    \hat H =\hat H_{DQD}+\hat H_\mathrm{leads}+\hat H_T+\hat H_{QPC}+\hat H_{DQD-QPC}.
\end{equation}
It consists of the Hamiltonian of the DQD
\begin{equation}
\hat H_{DQD} = E_L \hat a_L^{\dag}\hat a_L +E_R \hat a_R^{\dag} \hat a_R + \Omega(\hat a_L^{\dag}\hat a_R + \hat a_R^{\dag}\hat a_L) +U\hat n_L \hat n_R,
\nonumber
\end{equation}
where the operators $\hat a_L^{\dag}$ and $\hat a_R^{\dag}$ create an electron on the left and right quantum dot level with energy $E_L$ or $E_R$, respectively. The tunnel coupling between the levels is denoted as $\Omega$ and $\hat n_\alpha= \hat a_\alpha^{\dag}\hat a_\alpha=0,1$, $\alpha=L,R$ is the occupation number operator of each quantum dot. The inter-dot Coulomb interaction is denoted as $U$. The electrons in the leads are treated as non-interacting and are given by the Hamiltonian
\begin{equation}
\hat H_\mathrm{leads}=\sum_{k,\alpha=L,R} \epsilon_{k\alpha}\, \hat a_{k\alpha}^{\dag}\hat a_{k\alpha}^{\phantom{\dag}},
\end{equation}
where the operators $\hat a_{k\alpha}^{\dag}$ create an electron in lead $\alpha=L,R$ with momentum $k$ and energy $\epsilon_{k\alpha}$. The coupling between the DQD and the leads is accounted for by the standard Hamiltonian
\begin{equation}
\hat H_{T} = \sum_{k,\alpha=L,R} (t_{k\alpha}\hat a_{k\alpha}^{\dag}\hat a_{\alpha} + t_{k\alpha}^{\ast}\hat a_{\alpha}^{\dag}\hat a_{k\alpha}),
\end{equation}
which connects the left (right) lead to the left (right) quantum dot. Finally, the QPC is modeled as a tunnel barrier
\begin{equation}
\hat  H_{QPC}  = \sum_{k,\alpha=L,R}{\bar \epsilon_{k\alpha}}c_{k\alpha }^{\dag}c_{k\alpha }^{\dag}+\sum_{k,k'} (\bar t_{kk'}\hat c_{kL}^{\dag}\hat c_{k'R} + \bar t_{kk'}^{\ast}\hat c_{kR}^{\dag}\hat c_{k'L}),
\nonumber
\end{equation}
where the first sum corresponds to the electronic reservoirs on the left ($\alpha=L$) and right side ($\alpha=R$) of the QPC and the second sum describes the coupling of states in different leads with tunnel coupling $\bar t_{kk'}$.

If the QPC only couples to the total charge of the DQD, the charge detection is not expected to cause decoherence of the coherent oscillations of electrons inside the DQD. It is, however, interesting to investigate how an asymmetrically coupled QPC will affect the transport in the DQD. To this end, we assume that the QPC, besides the coupling to the total charge, has an additional capacitive coupling to the left quantum dot only. The charge occupation of the left quantum dot modulates the transparency of the QPC according to the Hamiltonian
\begin{equation}
\hat  H_{DQD-QPC} = \sum_{k,k'}\hat n_L (\delta\bar t_{kk'}\hat c_{kL}^{\dag}\hat c_{k'R}+\delta\bar t_{kk'}^*\hat c_{kR}^{\dag}\hat c_{k'L}).
\end{equation}
Here $\delta\bar t_{kk'}$ is the change of the QPC tunnel coupling in response to an (additional) electron occupying the left quantum dot.

We now follow Gurvitz in deriving a Markovian  GME for the reduced density matrix $\hat\rho_{DQD}$ of the DQD obtained by tracing out the electronic leads and the QPC. The details of the derivation can be found in Refs.~\cite{Gurvitz1996,Gurvitz1997}. We assume that a large bias is applied between the electronic leads, such that electron transport is uni-directional from the left to the right electrode. The electronic reservoirs have a continuous density of states and the discrete levels of the DQD are situated well inside the transport window. Under these assumptions, we may formulate a Markovian GME for the $n$-resolved reduced density matrix $\hat\rho_{DQD}(n,t)$. Its diagonal elements are the probabilities for the DQD to be either empty, having only left or right quantum dot occupied, or to be doubly-occupied, while $n$ electrons have been collected in the right lead during the measuring time $t$.

The diagonal elements of $\hat\rho_{DQD}(n,t)$ are denoted as $\rho_0(n,t)$, $\rho_L(n,t)$, $\rho_R(n,t)$, and $\rho_d(n,t)$. Additionally, we need the coherences between the left and the right quantum dot levels, denoted as $\rho_{LR}(n,t)$ and $\rho_{RL}(n,t)$. Coherences between states with different charge occupations are excluded. Since the off-diagonal elements fulfil $\rho_{RL}(n,t)=\rho_{LR}^*(n,t)$, it suffices to consider the real and imaginary parts of $\rho_{LR}(n,t)$. The elements of the reduced density matrix can then be collected in the vector
\begin{equation}\label{Eq:vecRho}
    |\rho(n,t)\rrangle\equiv[\rho_0,\rho_L,\rho_R,\Re[\rho_{LR}],\Im[\rho_{LR}],\rho_d]^T(n,t).
\end{equation}
The corresponding $z$-dependent reduced density matrix follows from the definition in Eq.~(\ref{eq:zdepRDM}) and reads
\begin{equation}
|\rho(z,t)\rrangle\equiv\sum_n|\rho(n,t)\rrangle z^n.
\end{equation}
The Markovian GME then takes the form
\begin{equation}\label{Eq:master_eq}
    \partial_t |\rho(z,t)\rrangle=\mathbf{M}(z)|\rho(z,t)\rrangle,
\end{equation}
with the rate matrix  reading
\begin{equation}
   \mathbf{M}(z)= \left[
      \begin{array}{cccccc}
        -\Gamma_L & 0 & z\,\Gamma_R & 0 & 0 & 0 \\
        \Gamma_L & 0 & 0 & 0 & -2\Omega & z\,\widetilde\Gamma_R \\
        0 & 0 & -\Gamma_R-\widetilde\Gamma_L & 0 & 2\Omega & 0 \\
        0 & 0 & 0 & -\Gamma & -\epsilon & 0 \\
        0 & \Omega & -\Omega & \epsilon & -\Gamma & 0 \\
        0 & 0 & \widetilde\Gamma_L & 0 & 0 & -\widetilde\Gamma_R \\
      \end{array}
    \right],
\end{equation}
where $\epsilon\equiv E_R-E_L$ is the energy detuning of the two quantum dot levels.
Additionally, the rate
\begin{equation}
\Gamma=\frac{1}{2}(\Gamma_R+\widetilde\Gamma_L+\gamma)
\end{equation}
determines the decay of the off-diagonal elements of $\hat\rho_{DQD}(n,t)$ and the broadening of the electronic levels. The electronic tunneling rates depend on the charge occupation of the DQD and read
\begin{equation}
\Gamma_{\alpha}=\frac{2\pi}{\hbar}\mathcal{D}_{\alpha}(E_{\alpha})|t_{k\alpha}|^2,\,\,\,\, \alpha=L,R
\end{equation}
and
\begin{equation}
\widetilde\Gamma_{\alpha}=\frac{2\pi}{\hbar}\mathcal{D}_{\alpha}(E_{\alpha}+U)|t_{k\alpha}|^2,\,\,\,\, \alpha=L,R,
\end{equation}
where $\mathcal{D}_{\alpha}$ denotes the density of states in lead $\alpha=L,R$, and the tunneling amplitudes $t_{k\alpha}$ are assumed to be $k$-independent. Here, $\Gamma_L$ is the tunneling rate from the left lead onto the left quantum dot, if the DQD is empty initially. On the other hand, if the right quantum dot is already occupied, electrons tunnel into the left quantum dot at rate $\widetilde\Gamma_L$. Similarly, electrons tunnel from the right quantum dot into the right lead with rate $\Gamma_R$, if the left quantum dot is empty, and with rate $\tilde\Gamma_R$, if the left quantum dot is occupied. Without inter-dot interactions, $U=0$, we have $\Gamma_{L(R)}=\widetilde\Gamma_{L(R)}$. Factors of $z$ have been included in the off-diagonal elements of $\mathbf{M}(z)$ corresponding to charge transfers from the right quantum dot to the right lead.

Finally, the decoherence rate introduced by the QPC is given by \cite{Gurvitz1997}
\begin{equation}\label{Eq:detector_decoherence_rate}
    \gamma=\frac{eV_d}{2\pi\hbar}(\sqrt{T}-\sqrt{\widetilde T}\,)^2,
\end{equation}
where $V_d$ is the bias applied accross the QPC. The transmission probability for electrons to tunnel through the QPC is
\begin{equation}
T=(2\pi)^2|\bar t_{k,k'}|^2\overline{\mathcal{D}}_L\overline{\mathcal{D}}_R,
\end{equation}
when the left quantum dot is empty. In contrast, when the left quantum dot is occupied, the QPC transmission reads
\begin{equation}
 \widetilde T=(2\pi)^2|\bar t_{k,k'}+\delta\bar t_{k,k'}|^2\overline{\mathcal{D}}_L\overline{\mathcal{D}}_R.
\end{equation}
Above, the symbols $\overline{\mathcal{D}}_{L(R)}$ denote the density of states in the left (right) lead of the QPC.

\section{Calculations}
\label{sec:calculations}

We now evaluate the FCS by formally solving Eq.~(\ref{Eq:master_eq}). We consider fluctuations in the stationary state, which we suppose has been reached at $t=0$, when we start counting charges. The stationary state is denoted as $|0\rrangle$ and is obtained by solving $M(z=1)|0\rrangle=0$ with the normalization condition $\llangle\tilde 0|0 \rrangle=1$, where $\llangle\tilde 0|=[1,1,1,0,0,1]$. From Eq.~(\ref{Eq:master_eq}), the GF can now be written as
\begin{equation}
\label{eq:GFintermsofM}
\mathcal{G}(z,t)=\llangle\tilde 0|e^{\mathbf{M}(z)t}|0\rrangle.
\end{equation}
It is easy to verify that this expression fulfils the condition $\mathcal{G}(z=1,t)=1$ for the GF.
It is a general, formally exact result, which yields the complete FCS at any time given the matrix $\mathbf{M}(z)$. However, in practice the expression may be difficult to evaluate due to the matrix exponentiation, in particular if the aim is to calculate the high (factorial) moments or (factorial) cumulants. In our recent work~\cite{Kambly2011}, we developed a simple method to evaluate the high-order, time-dependent statistics for these types of problems and we will also be using this method here. For details of the method, we refer the interested reader to Appendix A of Ref.~\cite{Kambly2011}.

\begin{figure}
\begin{center}
\includegraphics[width=0.45\textwidth, trim = 0 0 0 0, clip]{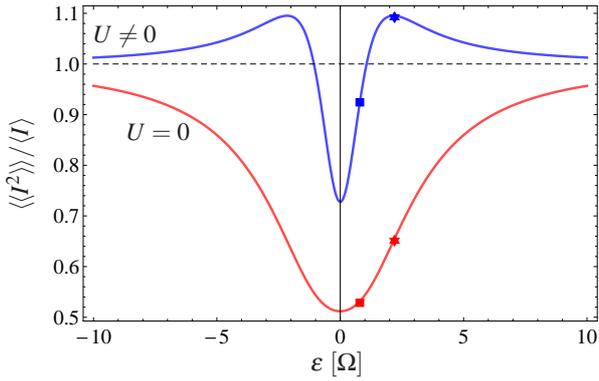}
\caption{Fano factor versus the energy detuning $\epsilon$. Results are shown with ($U\neq0$) and without ($U=0$) strong Coulomb interactions on the DQD. With strong Coulomb interactions, only 0 or 1 electrons can occupy the DQD. In contrast, without Coulomb interactions the DQD may also be doubly occupied. The QPC is coupled symmetrically to the DQD ($\gamma=0$). The other parameters are $\Gamma_R=\frac{1}{3}\Gamma_L$ and $\Omega=\hbar\Gamma_L$.  The squares and stars mark $\epsilon=0.8\Omega$ and $\epsilon=2.2\Omega$, respectively.}
\label{Fig:FanoFactor}
\end{center}
\end{figure}

In addition to the finite-time FCS, it is interesting to investigate the GF at long times. In this limit, the GF takes on a large-deviation form
\begin{equation}
\mathcal{G}(z,t)\propto e^{t\Theta(z)},
\end{equation}
where the rate of change is determined by the eigenvalue of $\mathbf{M}(z)$ with the largest real-part, i.~e.
\begin{equation}
\Theta(z)=\max_j\{\lambda_j(z)\}.
\end{equation}
From $\Theta(z)$ we may obtain the (factorial) CGF for the zero-frequency cumulants of the current. The zero-frequency cumulants of the current are
\begin{equation}\label{Eq:current_Cumu}
    \llangle I^m \rrangle=\partial_z^m\Theta(e^z)|_{z\rightarrow 0},
\end{equation}
while the corresponding factorial cumulants read
 \begin{equation}\label{Eq:current_factCumu}
    \llangle I^m \rrangle_F=\partial_z^m\Theta(z+1)|_{z\rightarrow 0}.
\end{equation}
In general, we can assume that the matrix $\mathbf{M}(z)$ at $z=1$ has a single eigenvalue equal to zero, i.~e.~$\lambda_0(1)=0$, corresponding to the (unique) stationary state, while all other eigenvalues have negative real-parts, ensuring relaxation toward the stationary state. For values of $z$ close to unity, we expect that $\lambda_0(z)$ develops adiabatically from 0 and still determines $\Theta(z)$ such that $\Theta(z)=\lambda_0(z)$ for $z\simeq1$. The derivatives of $\lambda_0(z)$ with respect to $z$ at $z=1$ then determines the (factorial) cumulants of the current according to Eqs.~(\ref{Eq:current_Cumu}) and (\ref{Eq:current_factCumu}).

\begin{figure}
\begin{center}
\includegraphics[width=0.95\columnwidth, trim = 0 0 0 0, clip]{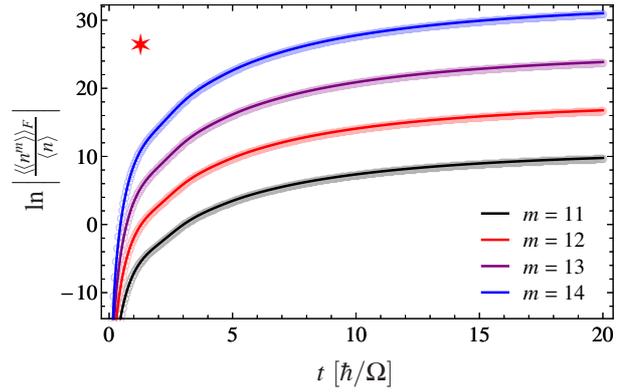}
\caption{Time-dependent factorial cumulants without interactions ($U=0$). The factorial cumulants $\llangle n^m\rrangle(t)$ of order $m=11$ through $m=14$ are shown as functions of time. The results correspond to the point marked with a red star in Fig.~\ref{Fig:FanoFactor}. The factorial cumulants do not oscillate, as expected without interactions. The full lines indicate numerical results, while circles show the approximation given by Eq.~(\ref{eq:largeFacGB}).}
\label{Fig:FCs_fcts_of_time_zeroU}
\end{center}
\end{figure}

Again, for large matrices $\mathbf{M}(z)$, it might not be viable to directly calculate the eigenvalue $\lambda_0(z)$ and its derivatives with respect to $z$ at $z=1$. This problem may be circumvented by considering the calculation of $\lambda_0(z)$ as a perturbation problem around $z=1$. The matrix $\mathbf{M}(z)$  is written as $\mathbf{M}(z)=\mathbf{M}(1)+\delta \mathbf{M}(z)$, where $\mathbf{M}(1)$ is the unperturbed matrix with eigenvalue $\lambda_0(z=1)=0$ and $\delta\mathbf{M}(z)=\mathbf{M}(z)-\mathbf{M}(1)$ is the perturbation. The eigenvalue $\lambda_0(z)$ can then be calculated order by order in $z$ using the recursive perturbation method developed in Refs.~\cite{Flindt2005,Flindt2008,Flindt2010}. This method yields the (factorial) cumulants of the current and will be used below.

\begin{figure*}
\begin{center}
\includegraphics[width=0.95\textwidth, trim = 0 0 0 0, clip]{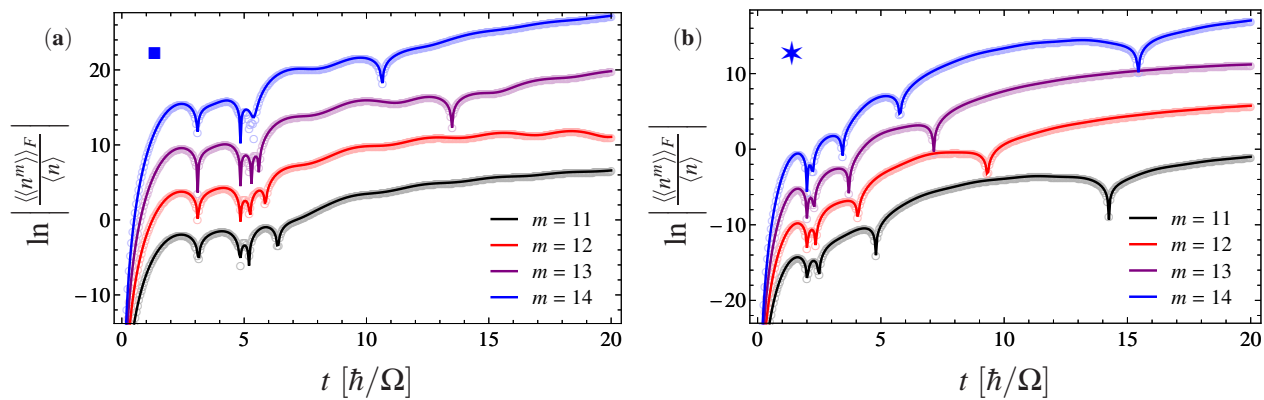}
\caption{Time-dependent factorial cumulants with strong Coulomb interactions.  The factorial cumulants $\llangle n^m\rrangle_F(t)$ of order $m=11$ through $m=14$ are shown as functions of time. The panels correspond to the points marked with a blue square (a) and a blue star (b) in Fig. \ref{Fig:FanoFactor}. The curves are shifted for clarity. We show the logarithm of the absolute value of the factorial cumulants such that downwards-pointing spikes correspond to a factorial cumulant going through zero. Full lines are numerical results, whereas empty circles correspond to the approximation given by Eq.~(\ref{Eq:approxFC_one_pair}).}
\label{Fig:FCs_fcts_of_time}
\end{center}
\end{figure*}

Finally, it is important to understand the connection between the FCS at finite times and in the long-time limit. As discussed in the previous section, the high (factorial) cumulants are related to the singularities of the (factorial) CGF in the complex plane of $z$. At finite times, the (factorial) CGF has logarithmic singularities at values of $z$, where the (factorial) MGF is zero. In contrast, in the long-time limit, the singularities of the (factorial) CGF are determined by the singularities of the eigenvalues of $\mathbf{M}(z)$. Typically, the eigenvalues have square-root branch points at the degeneracy points, where two eigenvalues are equal, i.~e.~$\lambda_0(z_c)=\lambda_1(z_c)$ for some $z_c$. Considering now the GF at finite times close to such a degeneracy point, we may approximate the GF in Eq.~(\ref{eq:GFintermsofM}) as
\begin{equation}
\mathcal{G}(z,t)=\sum_j c_j(z)e^{\lambda_j(z)t}\simeq c_0(z)e^{\lambda_0(z)t}+c_1(z)e^{\lambda_1(z)t},
\end{equation}
where the coefficients $c_j(z)$ depend on the initial condition and only the contributions from the two largest eigenvalues have been included. Solving for the zeros of $\mathcal{G}(z,t)$, we obtain the equations
\begin{equation}
\lambda_0(z)=\lambda_1(z)+\frac{\log \{ c_1(z)/c_0(z)\}+i\pi(2n+1)}{t},
\end{equation}
where $n$ is an integer. Importantly, we see that the second term on the right hand side vanishes in the limit $t\rightarrow\infty$. This analysis shows that the zeros of the GF as functions of time move towards the solutions of the equation $\lambda_0(z)=\lambda_1(z)$, which also determines the branch-point singularities in the long-time limit cf.~the discussion above. This connects the finite-time FCS with its long-time behavior.

\section{Results}
\label{sec:results}

We are now ready to illustrate the use of factorial cumulants on the concrete example of charge transport through a DQD. We analyze several different parameter regimes of the system which are discussed in turn. To begin with, it is instructive to consider the Fano factor $F$ of the transport in the long-time limit
\begin{equation}
F=\left.\frac{\llangle n^2\rrangle(t)}{\langle n\rangle(t)}\right|_{t\rightarrow\infty}=\frac{\llangle I^2\rrangle}{\langle I\rangle}
\end{equation}
given as the ratio of the zero-frequency current noise over the mean current. Figure~\ref{Fig:FanoFactor} shows the Fano factor as a function of the energy dealignment $\varepsilon$ without any decoherence due to the QPC, $\gamma=0$. We present results with ($U\neq 0$) and without interactions ($U=0$). For the non-interacting case, the Fano factor is never super-Poissonian ($F>1$) as expected for uni-directional transport with generalized binomial statistics. In contrast, for the interacting case there are certain ranges of the dealignment, where the noise becomes super-Poissonian. However, there is also a range of dealignments around $\varepsilon=0$, where the noise is sub-Poissonian ($F<1$), and in this regime a measurement of the Fano factor would not give any clear evidence of interactions. We note that recent noise measurements on transport through vertically coupled quantum dots are in qualitative agreement with the results shown for the interacting case \cite{Barthold2006,Kiesslich2007,Sanchez2008}.

We mark two points on the curves corresponding to values of the dealignment, where the noise in the interacting case is either sub-Poissonian (squares)  or super-Poissonian (stars). As we demonstrate now, the factorial cumulants give clear signatures of the interactions even in the cases with sub-Poissonian noise, where no conclusions can be drawn from the Fano factor alone. (We note that we also find oscillating factorial cumulants with symmetric rates $\Gamma_L=\Gamma_R$, where the noise is always sub-Poissonian.)

\begin{figure}
\begin{center}
\includegraphics[width=0.45\textwidth, trim = 0 0 0 0, clip]{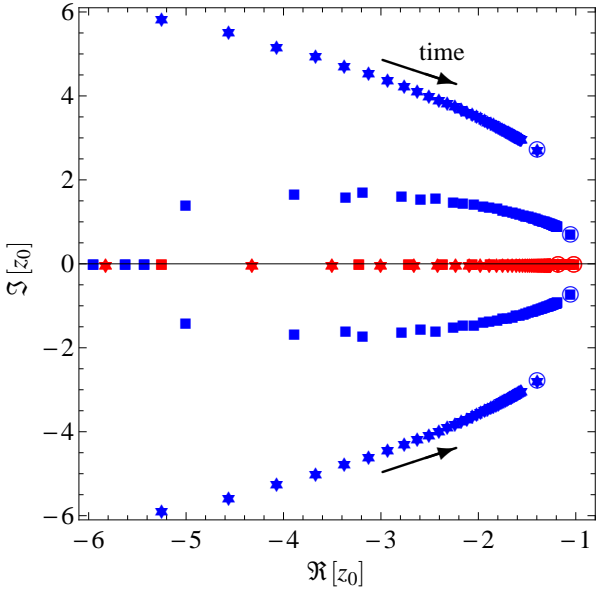}
\caption{Motion of the dominant singularities with time. The singularities have been extracted from the high order factorial cumulants in Figs.~\ref{Fig:FCs_fcts_of_time_zeroU}, \ref{Fig:FCs_fcts_of_time} and \ref{Fig:FCs_fcts_of_order}. The blue stars and squares are complex singularities corresponding to the oscillations of the factorial cumulants in Fig.~\ref{Fig:FCs_fcts_of_time}. The red stars and squares are real singularities corresponding to the non-interacting case considered e.~g.~in Fig.~\ref{Fig:FCs_fcts_of_time_zeroU}. The encircled points correspond to the long-time limits considered in Fig.~\ref{Fig:FCs_fcts_of_order}. The time is varied from $t_i=\hbar/\Omega$ to $t_f=50\hbar/\Omega$ in steps of $\Delta t=\hbar/\Omega$.}
\label{Fig:sngs_time_dep_complex_plane}
\end{center}
\end{figure}

In Fig.~\ref{Fig:FCs_fcts_of_time_zeroU} we show the time-dependent factorial cumulants for the non-interacting case, corresponding to the point marked with a star in Fig.~\ref{Fig:FanoFactor}. For the point marked with a square similar results are obtained. Without interactions, the FCS is generalized binomial and the factorial cumulants are expected to follow Eqs.~(\ref{Eq:GB_FCs}) and (\ref{eq:largeFacGB}), which predict no oscillations of the factorial cumulants as functions of time or any other parameter. This prediction is confirmed by Fig.~\ref{Fig:FCs_fcts_of_time_zeroU}, where a clearly monotonic behavior is found as a function of time. Moreover, from the calculated factorial cumulants, we may extract $p_\mathrm{max}$ in Eq.~(\ref{eq:largeFacGB}) as a function of time. Inserting $p_\mathrm{max}$ back into Eq.~(\ref{eq:largeFacGB}), we can compare this expression with the numerical results for the high-order factorial cumulants. In Fig.~\ref{Fig:FCs_fcts_of_time_zeroU}, the predicted behavior based on Eq.~(\ref{eq:largeFacGB}) is shown with circles and is seen to be in very good agreement with the full numerics.

Next, we turn to the time-dependent factorial cumulants in the interacting case. In Fig.~\ref{Fig:FCs_fcts_of_time} we show the high-order factorial cumulants corresponding to the two dealigments marked with stars and squares in Fig.~\ref{Fig:FanoFactor}. In this case, the factorial cumulants oscillate as functions of time in contrast to the non-interacting situation, where no oscillations are observed. To best visualize the oscillations, we show the logarithm of the absolute value of the factorial cumulants. Downwards-pointing spikes then correspond to the factorial cumulants passing through zero and changing sign. The oscillating factorial cumulants are a clear signature of interactions in the transport and they show that the FCS for this system is not generalized binomial, neither when the noise is sub-poissionian (square) nor super-poissionian (star).

Again, we can understand the high-order factorial cumulants using the expressions from Sec.~\ref{sec:highcumu}. In this case, when the FCS is not generalized binomial, the high-order factorial cumulants are expected to follow Eq.~(\ref{Eq:approxFC_one_pair}), which assumes that the factorial CGF has a complex-conjugate pair of singularities. At finite times, the factorial FCS has logarithmic singularities corresponding to the zeros of the factorial MGF. With only a single dominant pair of singularities, $z_0$ and $z_0^*$, the expression for the high-order factorial cumulants, Eq.~(\ref{Eq:approxFC_one_pair}) simplifies for finite times to~\cite{Flindt2009,Kambly2011}
\begin{equation}
\llangle n^m \rrangle_{(F)} \simeq -\frac{2(m-1)!}{\left|z_0\right|^{m}}\cos\left(m\arg\left[z_0\right]\right).
\label{Eq:HighFClog}
\end{equation}
From four consecutive high-order factorial cumulants, we can solve this relation for the dominant pair of singularities as functions of time using the methods described in Refs.~\cite{Zamastil2005,Flindt2010,Kambly2011} (see e.~g.~Appendix B of Ref.~\cite{Kambly2011}). Inserting the solution back into Eq.~(\ref{Eq:HighFClog}), we can benchmark the approximation against the numerical results. The approximation is shown with circles in Fig.~\ref{Fig:FCs_fcts_of_time} and is seen to be in excellent agreement with the full numerics.

Having extracted the dominant singularities as functions of time in the non-interacting and interacting cases, we may investigate their motion in the complex plane. In Fig.~\ref{Fig:sngs_time_dep_complex_plane} we show the dominant singularities as they move with time. In the non-interacting case ($U=0$), corresponding to the factorial cumulants shown in Fig.~\ref{Fig:FCs_fcts_of_time_zeroU}, the dominant singularity (marked with a red star) moves along the negative real-axis as expected for generalized binomial statistics. This behavior should be contrasted with the interacting case ($U>0$), corresponding to the factorial cumulants shown in Fig.~\ref{Fig:FCs_fcts_of_time}. In this case, the dominant singularities (marked with blue squares and stars) are no longer real and they now move in the complex plane as functions of time. We stress that this behavior cannot occur for a non-interacting system and should thus be taken as a signature of interactions.

\begin{figure}
\begin{center}
\includegraphics[width=0.95\columnwidth, trim = 0 0 0 0, clip]{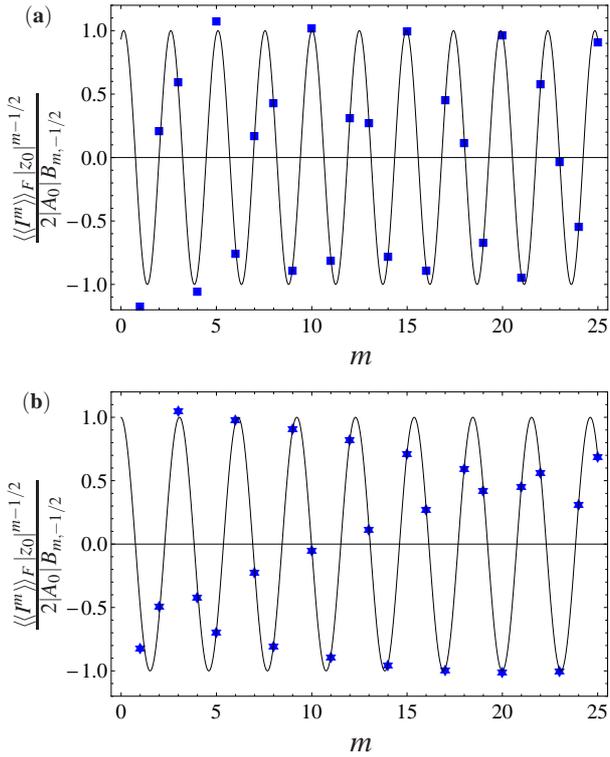}
\caption{Factorial cumulants $\langle\langle I^m \rangle\rangle_F$ of the current as functions of their order $m$. We compare numerical results (marked with symbols) and the approximation (full line) given by Eq.~(\ref{Eq:approxFC_one_pair}). Panel ({\bf a}) corresponds to the point marked with a blue square in Fig.~\ref{Fig:FanoFactor}. Panel ({\bf b}) corresponds to the point marked with a blue star. The corresponding dominant singularities extracted from the numerical results are encircled in Fig.~\ref{Fig:sngs_time_dep_complex_plane}.}
\label{Fig:FCs_fcts_of_order}
\end{center}
\end{figure}

In Fig.~\ref{Fig:sngs_time_dep_complex_plane}, we also indicate the points in the complex plane to which the dominant singularities move in the long-time limit (encircled points). As discussed in the previous section, these points correspond to the dominant singularities of the factorial CGF for the zero-frequency factorial cumulants of the current. We extract the position of these singularities by calculating the high order factorial cumulants of the current $\llangle I^m\rrangle_F$ using the recursive scheme developed in Ref.~\cite{Flindt2010}, here adapted to the calculation of factorial cumulants. The results for the factorial cumulants of the current are shown in Fig.~\ref{Fig:FCs_fcts_of_order} as functions of the order $m$. Together with the numerical results, we show the approximation in Eq.~(\ref{Eq:approxFC_one_pair}) with full lines. From four consecutive high-order factorial cumulants we have extracted the parameters entering Eq.~(\ref{Eq:approxFC_one_pair}) using the method proposed in Ref.~\cite{Flindt2010}. Typically, in the long-time limit, the singularities are square-root branch points such that $\mu_j=-1/2$ in Eq.~(\ref{Eq:approxFC_one_pair}). Figure~\ref{Fig:FCs_fcts_of_order} shows that Eq.~(\ref{Eq:approxFC_one_pair}) provides an excellent approximation of the numerical results and it allows us to extract the dominant singularities in the long-time limit (encircled points) in Fig.~\ref{Fig:sngs_time_dep_complex_plane}. As anticipated, the dominant singularities at finite times move towards the long-time singularities. In the long-time limit, the singularities cease to move with time and the high-order (factorial) cumulants will no longer oscillate as function of time (but still as functions of other parameters).

\begin{figure}
\begin{center}
\includegraphics[width=0.95\columnwidth, trim = 0 0 0 0, clip]{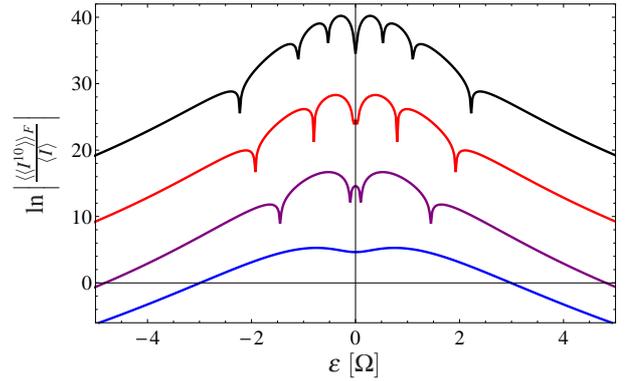}
\caption{Factorial current cumulants with detector-induced dephasing. The parameters are $\Omega=\hbar\Gamma_L$ and $\Gamma_R=(1/3)\Gamma_L$ and $\gamma=0$ (black), $0.1\Omega/\hbar$ (red), $0.2\Omega/\hbar$ (purple), $0.4\Omega/\hbar$ (blue). The curves are shifted for clarity.}
\label{Fig:FC10withDetector}
\end{center}
\end{figure}

Finally, we turn to the influence of detector-induced dephasing. In Fig.~\ref{Fig:FC10withDetector} we consider the situation where the QPC charge detector is asymmetrically coupled to the two quantum dots, thereby causing dephasing of electrons passing through the DQD. Due to strong Coulomb interactions, the DQD can only be either empty or occupied by one (additional) electron at a time. Without detector-dephasing ($\gamma=0$), clear oscillations of the tenth factorial cumulant of the current are observed as a function of the energy dealignment $\varepsilon$. However, as the dephasing rate is increased, the oscillations are gradually washed out and they essentially vanish in the strong dephasing limit with $\gamma=0.4\Omega/\hbar$. Thus, in this case, dephasing of the coherent oscillations of electrons inside the DQD seems to reduce the signatures of interactions in the FCS.

\section{Conclusions}
\label{sec:conclusions}

We have discussed our recent proposal to detect interactions among electrons passing through a nano-scale conductor by measuring time-dependent high-order factorial cumulants. For non-interacting electrons in a two-terminal scattering problem, the full counting statistics is always generalized binomial, the zeros of the generating function are real and negative, and consequently the factorial cumulants do not oscillate as functions of the observation time or any other system parameter. In contrast, oscillating factorial cumulants must be due to interactions in the charge transport. In cases where the factorial cumulants oscillate, the zeros of the generating function have moved away from the real-axis and into the complex plane. As we have shown, the motion of the dominant zeros of the generating function can be deduced from the oscillations of the high order factorial cumulants.

Here, we have illustrated these ideas with a system consisting of two coherently coupled quantum dots attached to voltage-biased electronic leads. The dynamics of the DQD was described using a Markovian generalized master equation which allowed us to treat strong coupling to the leads together with the coherent evolution of electrons inside the DQD. Interestingly, we found that even in cases where the Fano factor of the transport is sub-Poissonian, the high-order factorial cumulants still enable us to detect interactions among the charges passing through the DQD. Finally, we discussed the influence of detector-induced dephasing on the FCS and found that, for this model, such dephasing processes may reduce the oscillations of the high order factorial cumulants.

Our work leaves several open questions for future research. It is still not clear exactly under what conditions interactions cause oscillations of the factorial cumulants. This will require further careful investigations of the singularities of generating functions in FCS, for example as in the recent work on singularities in  FCS for molecular junctions~\cite{Utsumi2013}. The answer to this question may moreover come from future measurements of oscillating factorial cumulants in interacting nano-scale conductors. In this work, we have focused on Markovian master equations, and it would be interesting to investigate similar phenomena for non-Markovian systems \cite{Braggio2006,Flindt2008,Flindt2010,Marcos2011,Emary2011}.  Finally, a new and promising direction of research combines the zeros of generating functions and high order statistics with dynamical phase transitions in stochastic many-body systems~\cite{Ivanov2010,Garrahan2010,Flindt2013}.

\begin{acknowledgements}
We acknowledge many useful discussions with Markus B\"{u}ttiker. The work was supported by the Swiss NSF.
\end{acknowledgements}

\bibliographystyle{spphys}       % APS-like style for physics

\end{document}